# From Early Theories of Dzyaloshinskii-Moriya Interactions in Metallic Systems to Today's Novel Roads

Albert Fert[1], Mairbek Chshiev[2,3], André Thiaville[4], Hongxin Yang[5*]

[1]Unité Mixte de Physique, CNRS, Thales, Univ. Paris-Saclay, Université Paris-Saclay, Palaiseau 91767, France.

[2]Univ. Grenoble Alpes, CEA, CNRS, Spintec, Grenoble 38000, France.
[3]Institut Universitaire de France (IUF), Paris 75231, France.
[4]Laboratoire de Physique des Solides, Université Paris-Saclay, CNRS, Orsay 91405, France.
[5] National Laboratory of Solid State Microstructures, School of Physics, Collaborative Innovation Center of Advanced Microstructures, Nanjing University, Nanjing 210093, China

*Correspondence: hongxin.yang@nju.edu.cn

## Abstract

Since the early 1960's, the discovery of Dzyaloshinskii-Moriya interaction (DMI) helped to explain the physical mechanisms behind certain magnetic phenomena, such as net moment in antiferromagnets, or enhanced anisotropy field from heavy metals impurity in dilute Cu:Mn alloy. Since the researchers unveil the key role that DMI plays in stabilizing chiral Néel type magnetic domain wall and magnetic skyrmions, the studies on DMI have received growing interest. Governed by spin-orbit coupling (SOC) and various types of inversion symmetry breaking (ISB) in magnetic systems, DMI drives the forming of distinct morphologies of magnetic skyrmions. Our aim is to briefly introduce the research history of DMI and its significance in the field of modern spintronics.

## 1. Introduction

The Dzyaloshinskii-Moriya interaction (DMI) is an anti-symmetric interaction that forces the spins of neighboring atomic sites to align perpendicular to each other. The Heisenberg interaction between two spins favors parallel (ferromagnetic) or anti-parallel (antiferromagnetic) states, whereas DMI induces a clockwise or counter-clockwise rotation between the spins. The presence of DMI requires breaking inversion symmetry, and the existence of sizable spin-orbit coupling (SOC). It acts as a key ingredient for noncollinear magnetism and chiral magnetism, leading to chiral domain walls and magnetic skyrmions. Such peculiar spin textures are of great interest in both fundamental and application aspects. Racetrack memory and logic devices based on skyrmions and chiral domain walls are very promising spintronic candidates. From the non-centrosymmetric bulk magnets to the metallic multilayer systems, the DMI effect has been intensively studied both theoretically and experimentally.

Here, we recall the history of the early models of DMI, the density functional theory (DFT) approaches of calculating DMI, the discovery of magnetic skyrmions and chiral domain walls and their potential applications. Furthermore, we briefly discuss some prospects of interlayer DMI and possibilities of its ferroelectric control, including in systems comprising two-dimensional magnets.

## 2. Early models of the Dzyaloshinskii-Moriya interaction

Early developments of DMI were first aimed to explain the origin of "weak" ferromagnetism in some antiferromagnetic materials. Previously, it had been noted that some materials considered to be antiferromagnetic, such as $\alpha$-$Fe_2O_3$, or the $MnCO_3$ and $CoCO_3$ carbonate compounds, exhibited spontaneous magnetization behavior, with very small magnetic moment compared to that of respective magnetic atoms. Néel attributed the net moment in these antiferromagnets to an impurity effect.[1] Thus, the purity and uniformity of these crystals would strongly affect the ferromagnetic properties, and in an ideal antiferromagnetic crystal, such spontaneous magnetization would vanish. However, later reports showed that ferromagnetism could persist in very pure crystals.[2] Meanwhile, Li proposed that the net moment in these crystals could originate from canted spins in the antiferromagnetic domain walls, as the spin canting could possibly give rise to the net moment in the crystals.[3] However, the formation of such domain walls is not energetically favorable. In 1957, Dzyaloshinskii used the Landau second-order phase transition theory in order to demonstrate that "weak" ferromagnetism in $\alpha$-$Fe_2O_3$ could be due to the spin canting state of the material.[4,5] Specifically, the symmetry of a magnetic crystal is determined by the space group of atom and spin distributions, leading to different classes of magnetic states. As shown in Fig. 1(a), three magnetic states can be identified in $\alpha$-$Fe_2O_3$, namely, state I with spins directed along the crystal axis, state II with spins lying in one of the planes of symmetry, and state III with some spins along second-order axes. The state I is the commensurate antiferromagnetic state without net moment, while states II and III can exhibit spontaneous magnetic moment. By investigating the thermodynamic potentials of α-$Fe_2O_3$, Dzyaloshinskii proved that the transition between states I and II (or III) will occur at a given temperature and pressure. He also suggested that the presence of a different type of spin interaction was responsible for the aforementioned transition that causes a tilting between the spins of neighboring atomic sites.

At the beginning of 1960, Moriya pointed out that the spin interaction Dzyaloshinskii suggested should be an anti-symmetric interaction of the form:

$$\vec{D} \cdot [\vec{S_i} \times \vec{S_j}] \tag{1}$$

with $\vec{D}$ named the Dzyaloshinskii-Moriya interaction (DMI) vector, $\vec{S_i}$ and $\vec{S_j}$ indicate the spins of two atomic sites $i$ and $j$.[6] By including the effect of spin-orbit coupling to Anderson's superexchange theory,[7] Moriya deduced that the presence of $\vec{D}$ requires spin-orbit coupling and inversion symmetry breaking in magnetic crystals.

Several months later, Moriya developed a general theory to describe the microscopic mechanism of DMI.[8] In Moriya's model, for two magnetic atoms with only $3d$ orbitals at atomic sites $i$ and $j$, the two-site Hubbard-type Hamiltonian reads:

$$H = H_0^i + H_0^j + T^{ij} + H_{SO}^i + H_{SO}^j, \tag{2}$$

where $H_0^i$ denotes the localized $3d$ electrons on site $i$:

$$H_0^i = \sum_{m,\sigma} \varepsilon_{i,m} c_{im\sigma}^\dagger c_{im\sigma} + U \sum_{m\sigma \neq m'\sigma'} n_{im\sigma} n_{im'\sigma'}, \tag{3}$$

in which $\varepsilon_{i,m}$ is the orbital energy of $3d$ electron, and $U$ indicates the Coulomb repulsion.[8,9] The hopping between sites $i$ and $j$ reads:

$$T^{ij} = \sum_{n,m,\sigma} t_{in,jm} (c_{in\sigma}^\dagger c_{jm\sigma} + c_{jm\sigma}^\dagger c_{in\sigma}), \tag{4}$$

where $t_{ij}$ denotes the hopping integral between sites $i$ and $j$, and $t_{in,jm}$ is the contribution between orbitals $n$ and $m$ for $t_{ij}$, respectively. $H_{SO}^i$ in Eq. 2 represents the spin-orbit coupling (SOC) term at site $i$:

$$H_{SO}^i = \xi \boldsymbol{L_i} \cdot \boldsymbol{S_i}, \tag{5}$$

where $\boldsymbol{L_i}$ and $\boldsymbol{S_i}$ denote the angular momentum and spin momentum at atomic site $i$. Under the limit of large $U$ ($U >> t_{ij}$), the last three terms of Hamiltonian in Eq. (2) can be treated as perturbation to $H_0 = H_0^i + H_0^j$, the effective interaction between two atomic sites spins $\mathbf{S}_i$ and $\mathbf{S}_j$ can be derived as:

$$H_{eff} = -J_{ij}\mathbf{S}_i \cdot \mathbf{S}_j + \mathbf{D}_{ij} \cdot (\mathbf{S}_i \times \mathbf{S}_j) + \mathbf{S}_i : \mathbf{\Gamma}_{ij} : \mathbf{S}_j, \tag{6}$$

Here, the scalar $J_{ij}$ in the first term is the Heisenberg exchange interaction obtained from second order of perturbation of Hamiltonian $H_0$, which is a symmetric interaction with $J_{ij} = J_{ji}$ and has the order $(t_{ij})^2/U$. The second term is the DMI vector $\mathbf{D}_{ij}$, that can be obtained considering $H_{SO}$. $\mathbf{D}_{ij}$ is antisymmetric, with $\mathbf{D}_{ij} = -\mathbf{D}_{ji}$. The strength of $\mathbf{D}_{ij}$ is proportional to $\xi(t_{ij})^2/U$. When the fourth order perturbation is included, the symmetric tensorial interaction $\mathbf{\Gamma}_{ij}$ can be derived. $\mathbf{\Gamma}_{ij}$ has the smallest energy scale of $\xi^2(t_{ij})^2/U$, which can be neglected. A physical picture of this result can be described as follows: The electron hopping between nearest neighbor magnetic atoms does not occur with spin-flipping in the absence of SOC, and the neighboring spins prefer collinear configuration due to superexchange interactions. The spin-flipping hopping process of electrons between nearest neighbor magnetic atoms only occurs while SOC effect is considered. Such two sites spin-flipping hopping process defines the microscopic origin of DMI.

As a blueprint for the effect of crystal symmetry on DMI, Moriya proposed five criteria, later known as the Moriya rules.[8] If two magnetic ions located at the points *A* and *B*, respectively, and the center at *AB* is denoted by *C*, then:

1. When an inversion center located at *C*,

    $\vec{D} = 0$

2. When a mirror plane perpendicular to *AB* passes through *C*,

    $\vec{D}$ ∥ mirror plane or $\vec{D}$ ⊥ *AB*

3. When there is a mirror plane including *A* and *B*,

    $\vec{D}$ ⊥ mirror plane

4. When a two-fold rotation axis perpendicular to *AB* passes through *C*,

    $\vec{D}$ ⊥ two-fold rotation axis

5. When there is an n-fold axis (n ≥ 2) along *AB*,

$$\vec{D} \parallel AB$$

The schematic representations of Moriya rules are shown in Fig. 1(b). For a ferromagnetic state, the adjacent spins are parallel to each other. As shown in the left panel of Fig. 1(c), DMI vectors with opposite signs result in clockwise and anticlockwise rotation between ferromagnetic aligned spins.

In 1976, Smith predicted that for ferromagnetic metals, spin-orbit scattering of the conduction electrons by the nonmagnetic impurities could give rise to additional term of DMI arising from Ruderman-Kittel-Kasuya-Yoshida (RKKY) mechanism.[10,11] Fert and Lévy extended this theory, and, for the non-centrosymmetric situation shown in the right panel of Fig. 1(c), calculated the DMI arising from electron exchange scattering on the two magnetic atoms and SOC scattering on a non-magnetic atom with strong SOC. [12,13] They successfully explained the drastically enhanced anisotropy field induced by heavy $d$ metal (Au, Pt) impurities in Cu:Mn dilute alloys hosting spin glass states. The DMI vector of the model proposed by Fert and Lévy can be written as:

$$\vec{D}_{ijl}(\vec{R}_{li}, \vec{R}_{lj}, \vec{R}_{ij}) = -V_1 \frac{\sin[k_F(|\vec{R}_{li}|+|\vec{R}_{lj}|+|\vec{R}_{ij}|)+(\pi/10)Z_d](\vec{R}_{li}\cdot\vec{R}_{lj})(\vec{R}_{li}\times\vec{R}_{lj})}{|\vec{R}_{li}|^3|\vec{R}_{lj}|^3|\vec{R}_{ij}|}, \qquad (7)$$

where $\vec{R}_{li}$, $\vec{R}_{lj}$ and $\vec{R}_{ij}$ are the distance vectors of the three sides of the triangle formed by the magnetic ions at site $i$, $j$, and the spin-orbit center $l$. The parameter $V_1 = [135\pi\lambda_d\Gamma^2(\sin(Z_d\pi/10))/(32k_F^3 E_F^2)]$ refers to parameters of the electron gas ($k_F$, $E_F$), their exchange interaction with the magnetic atoms ($\Gamma$), and parameters of the $d$ electrons of the heavy metal impurity ($\lambda_d$, and $Z_d$).

The DMI mechanism of Fig. 1(c) has been extended by Fert to the non-centrosymmetric situation at an interface between a magnetic metal and a nonmagnetic metal (NM) of large SOC, [13] which leads to interfacial DMIs with, in most case, DMI vectors in the plane of the interface.

## 3. Spin spirals from interfacial Dzyaloshinskii–Moriya interactions

In the early 2000s, researchers unveiled non-collinear magnetic states in the 3d metal monolayer/NM heterostructures.[14-18] Magnetic frustration, i.e., competing antiferromagnetic Heisenberg exchange from further neighbors could give rise to non-collinear magnetic ground state in a magnet.[18,19] The energy spectrum of the spin spirals could serve as a describer for non-collinear magnetism, in which the spin moment at site $r_i$ can be generally described as $\hat{S}_i = [\cos(q \cdot r_i)\sin\vartheta, \sin(q \cdot r_i)\sin\vartheta, \cos\vartheta]$, where $q$ is the spiral wave vector, and $\vartheta$ denotes the cone angle. Figs. 2(a) plots four types of homogenous spin spirals, namely, cone Néel type, plane Néel type, cone Bloch type and plane Bloch type spirals.[17] Such collective rotation of spins can be considered as a generalized translation action from the point of view of the generalized Bloch theorem (gBT).[20,21]

Specifically, in a magnetic crystal, the eigenfunction $\Psi_k(r)$ for one-electron Hamiltonian $H$ takes the form of a Bloch function:

$$\Psi_k(r) = e^{-ik \cdot r} u_k(\mathbf{r}), \qquad (8)$$

where $k$ and $r$ represents momentum and position vectors, respectively, $u_k(\mathbf{r})$ is a periodic function with the same periodicity as the magnetic crystal, e is the Euler number and i denotes the imaginary unit. For a non-colinear magnetic periodical system, Sandratskii adopted the concept of spin space group (SSG) to depict the collective rotation actions of atomic spins.[20,21] The group element of SSG are the rotation actions, noted as $s_R$. Due to group isomorphism between SSG and generalized translation group, $s_R$ could be represented by generalized translation action $t_n$, with the latter is the element of generalized translation group. Because of the similarity between generalized translation action and ordinary translation action in periodical atomic systems, one can possibly associate generalized translation to momentum vector $k$ of Brillouin zone. The generalized translation operator $R_n$ is defined as:

$$R_n = e^{-iq \cdot t_n}, \qquad (9)$$

where $q$ is the aforementioned spiral wave vector. The rotation angles of atomic spins can be given by $q \cdot t_n$. In the generalized form of Bloch function $\Psi_k^q(r) = e^{-ik \cdot r} u_k^q(\mathbf{r})$, the spinors function $u_k^q(\mathbf{r})$ have the generalized periodicity of Hamiltonian, with $R_n\, u_k^q(\mathbf{r}) = u_k^q(\mathbf{r})$.

The gBT here offers the possibility of calculating the energy dispersion $E[q]$ associated with spin spiral vector length $q$, which can be applied in density functional theory (DFT), the Korringa–Kohn–Rostoker (KKR) frameworks and tight-binding method.[20] For the magnetic frustration-induced spin spiral ground states, $E[q] = E[-q]$, thus clockwise and anticlockwise rotating spin spirals are energetically degenerate, as plotted in Figs. 2(b). The presence of interfacial DMI (iDMI) will inevitably lift the degeneracy of two spin spirals with opposite chirality. From 2007 on, the Hamburg+Jülich group discovered long period spin spirals with a unique chirality in Mn monolayer on W (110) and W (001) substrates, which were the first experimental evidences of iDMI. [22,23]

Computational derivation of iDMI parameters needs to deal with SOC and non-collinear magnetism simultaneously. However, for a given angular momentum $l$ and spin momentum $s$ of a magnetic crystal, the SOC operator $l \cdot s$ cannot commute with the generalized translation operator $R_n$, thus SOC and gBT are exclusive. By consequence, the SOC effect cannot be included directly while calculating the spin spiral energy.

The Jülich group suggested a method to calculate the SOC-affected spin spiral energy dispersion by treating the SOC effect as a first-order perturbation.[24] For the ferromagnetic (FM)/heavy metal interfaces, one can introduce a plane Néel type spiral with $\hat{S}_i = [\cos(q \cdot r_i), \sin(q \cdot r_i), 0]$ to investigate the interfacial DMI. The Hamiltonian with SOC term $H_{SOC}$ included for the spin spiral reads:

$$H_{tot} = H_0 + H_{SOC}, \qquad (10)$$

where $H_0$ denotes the unperturbed spin spiral Hamiltonian. The Kohn-Sham equation of $H_0$ and $H_{tot}$ can be described as

$$H_0 \varphi_{0,v}(q) = \epsilon_{0,v}(q) \varphi_{0,v}(q), \qquad (11)$$

$$H_{tot} \varphi_{ft,v}(q,k) = (H_0 + H_{soc}) \varphi_{ft,v}(q,k) = \epsilon_{ft,v}(q,k) \varphi_{ft,v}(q,k), \qquad (12)$$

where $\varphi_{0,v}(q)$ and $\epsilon_{0,v}(q)$ are the unperturbed eigenstates and energy spectrum, respectively. $\epsilon_{ft,v}(q)$ is the spectrum of the total Hamiltonian $H_{tot}$ with eigenstates $\varphi_{ft,v}(q)$. By applying the magnetic force theorem, [25-27] the energy shift resulting from SOC effect $E_{DM}(q)$ can be obtained by summation over all occupied states:

$$E_{DM}(q) = \sum_v^{o.c.} \epsilon_{ft,v} - \sum_v^{o.c.} \epsilon_{0,v} \approx \sum_v n_v(q) \delta \epsilon_v(q), \qquad (13)$$

where $\delta \epsilon_v(q) = \langle \varphi_{0,v}(q) | H_{soc} | \varphi_{0,v}(q) \rangle$, and $n_v$ is the occupation number of the unperturbed states.

Here, as representative examples, Figs. 2 (c) -(d) plot the homogenous plane spin spiral in ferromagnetic Ir(111)/Fe/Pd and antiferromagnetic Rh(001)/Ir/Fe films.[28,29] When SOC is neglected, the energy minimum of spin spiral energy $E[q]$ locates at $q = 0$ (ferromagnetic ground state (Fig. 2(c)) or $q = \sqrt{2}/2$ (antiferromagnetic ground state, see Fig. 2(d)). Once SOC is included, the energy dispersions $E[q]$ for both cases in Figs. 2 (c)-(d) show an asymmetric behavior due to the presence of DMI. The DMI energy defined as $\Delta E_{DMI} = (E[q] - E[-q])/2$ shows a linear dependence on $q$, thus one can determine the effective DMI parameter using $D = \frac{dE[q]}{dq}$.

In 2017, Sandratskii proved that if the spin-orbit operator $\boldsymbol{l} \cdot \boldsymbol{s}$ is restricted to the direction of rotation axis $\hat{\boldsymbol{n}}$, the form of $(\boldsymbol{l} \cdot \hat{\boldsymbol{n}}) \cdot (\boldsymbol{s} \cdot \hat{\boldsymbol{n}})$ is commute with the generalized translation operator $\boldsymbol{R}_n$.[30] In another word, for a given spin spiral, if the SOC Hamiltonian $H_{SOC}$ in Eq. (10) is constrained to a single component along the direction of the rotational axis, the SOC included spin spiral energy spectrum of $H_{tot}$ can be obtained using self-consistent calculations. This approach is the so-called qSO method, which is an extension of the first-order perturbation theory. With the qSO method, the first-order perturbation theorem of gBT is no longer limited to the full-potential DFT software.[31-33]

## 4. Interface-induced Néel-type domain walls

Magnetic domains arise to minimize the sample's magnetostatic energy, given its shape. In between these domains, domain walls (DW) appear, whose structure and energy are the result of a trade-off between various energy terms.[34,35] The DWs can be classified into two types: the Bloch-type DWs and the Néel-type DWs. Within the former, magnetization rotates in a plane parallel to the wall plane, so that no magnetostatic volume charges appear, while within the latter the magnetization rotates in a plane perpendicular to the wall plane. The long-range dipole-dipole interaction dominates in bulk magnets and thicker magnetic films, thus the DW in such samples are usually of the Bloch-type, as shown in Fig. 3(a). In nanoscale samples such as ultrathin films and nanowires, the magneto-static energy is weakened, whereas the SOC effects like perpendicular magnetic anisotropy (PMA) and interface-induced DMI (iDMI) are enhanced due to the inversion symmetry breaking (ISB). Thus, the Néel-type DWs can be stabilized in thin films with PMA and sizeable iDMI. To distinguish the Néel DWs found in in-plane anisotropy thin films, the Néel-type DWs induced by iDMI are called Dzyaloshinskii DWs or chiral Néel DWs, as plotted in Fig. 3(a).[36]

The presence of chiral Néel DWs was confirmed by measuring the current-driven DW motion in Co/Pt thin films, in which the DWs velocity shows asymmetric aspect depending on the chirality of iDMI.[37,38] The direct experimental observation of chiral Néel DWs was reported by Chen et al. in Fe/Ni bilayers epitaxially grown on Cu(100) substrate,[39,40] as shown in Fig. 3(b). They confirmed that the growth order of Fe/Ni thin films allows determining the chirality of chiral Néel DWs that is caused by the iDMI at the Fe/Ni interface. Although in some cases such as tetragonal ferrimagnetic layers, Néel-type DWs may be induced by the bulk DMI, the presence of iDMI is the crucial ingredient for Néel-type DWs in perpendicularly magnetized ultrathin films.[41-44]

The interface-induced chiral Néel DWs are more advantageous compared to the Bloch-type DWs in current-driven dynamics. Both the Bloch-type DWs and chiral Néel DWs can be driven by spin-transfer torque (STT), while only the chiral Néel DWs can be driven by the spin-orbit torque (SOT), with higher efficiency (see Fig. 3(c)). [45-51] Parkin et al moreover proved that in synthetic antiferromagnetic (SAF), the velocity of chiral Néel DWs can be enhanced to several 100 m/s, which shows the potential of designing high-speed spintronic devices.[52,53]

## 5. Chirality-dependent total energy difference calculation of iDMI

Due to interfacial inversion symmetry breaking, the iDMI is ubiquitous in the ferromagnetic/nonmagnetic (FM/NM) interfaces. The chirality and magnitude of iDMI strongly depends on material combination and film thickness. Thus, optimizing the construction of magnetic multilayers is crucial for large iDMI. However, most theoretical studies using the first-order perturbation theorem of gBT focus on the properties of FM monolayer on the HM substrates. In 2015, Yang et al. developed the chirality-dependent total energy difference approach to determine the iDMI parameters. [54] As a representative example of the Co/Pt systems in Fig. 3(d), the microscopic DMI parameter $d_{tot}$ can be obtained by calculating the difference of the DFT energies $E_{cw}$ and $E_{acw}$ of clockwise and anticlockwise spin configurations, respectively, which reads:

$$d_{tot} = \frac{E_{cw} - E_{acw}}{m}, \qquad (14)$$

where $m$ depends on the spin spiral period, with additional possibility to evaluate DMI strength $d^k$ concentrated in a single atomic layer $k$. With this approach, the size, chirality and the energy sources of iDMI at the interfaces of various FM/HM heterostructures, FM/graphene and FM/oxide interfaces have been determined.[54-56] In particular, unlike the Fert-Lévy mechanism more suitable for FM/HM heterostructures, FM/2D and FM/oxide interfaces have been determined can be attributed to a more complex mechanism: the interface-induced twofold spin energy spectrum degeneracy breaking, known as the Rashba effect.[55-59] The Rashba Hamiltonian is described as

$$H_R = \alpha_R(\boldsymbol{\sigma} \times \boldsymbol{k}) \cdot \hat{\boldsymbol{z}}, \tag{15}$$

where $\alpha_R$ indicates the Rashba coefficient, $\boldsymbol{\sigma}$ is the Pauli matrix vector of atomic spins and $\boldsymbol{k}$ is the momentum of atomic orbitals. From theoretical models, several groups suggested that DMI could be induced by Rashba effect.[60-62] The relation between DMI strength $d$ and Rashba coefficient is described as:

$$d = 2\, k_R A, \tag{16}$$

where $k_R = \frac{2\alpha_R m_e}{\hbar^2}$ is a constant determined by Rashba coefficient $\alpha_R$, effective mass of electron $m_e$ and the reduced Planck constant $\hbar$, $A$ is the spin stiffness parameter. The energy source of Rashba-type DMI is contributed by the interfacial magnetic atoms rather than the adjacent non-magnetic atoms.[31,54,55] Moreover, for FM heterostructures with multiple interfaces, the chirality-dependent total energy difference approach allows extracting the iDMI contribution from each monolayer and interface, which can provide guidelines to maximize iDMI for FM multilayers.[56]

## 6. Magnetic Skyrmions

Skyrmions are particles-like swirling configurations. The concept of skyrmions was first proposed by Skyrme in 1962 when he tried to explain how subatomic particles can exist as discrete entities surrounded by a continuous nuclear field.[63] In 1975, Belavin and Polyakov proved that such metastable quasi-particles could exist in 2D ferromagnets.[64] From the 1990s, Bogdanov *et al.* theoretically predicted that magnetic skyrmions could be induced and stabilized by DMI.[65-67]

In 2009, magnetic skyrmions were discovered in the MnSi crystals by the Pfleiderer, Böni and colleagues, using small angle neutron scattering.[68,69] Shortly after, Yu *et al*. obtained the first real space images of skyrmions in $Fe_{0.5}Co_{0.5}Si$ films, by Lorentz transmission electron microscopy, as shown in Fig. 4(a).[70] For these cubic compounds of the B20 cristallographic type, the skyrmions are the Bloch type due to bulk DMI. The first instance of a skyrmions lattice in an ultrathin film was found in the Fe monolayer on Ir(110) substrate by Heinze *et al*. (see Fig. 4(b)).[71] In such monolayer, skyrmions are stabilized by iDMI and four-spins interactions. Later reports showed that isolated skyrmions could be found in Ir(111)/Fe/Pd ultrathin films due to iDMI [72]. Till today, bulk materials hosting skyrmions consist of a variety of acentric magnetic crystals. [73-78] In these magnets, the helicity of skyrmions varies from Bloch-type, Néel-type and antiskyrmions (see Fig. 4(c)) depending on the respective DMI vectors.

In the past decades, FM multilayer thin films hosting skyrmions have received greater attention due to their compatibility with the contemporary magnetic storage media and technologies. Moreover, by adjusting film thickness and material combinations in the FM based multilayers, one can elaborately control the iDMI, perpendicular magnetic anisotropy and exchange stiffness, thereby tuning the size, temperature stability and dynamics of skyrmions. In 2015, Chen *et al*. carefully tuned the interlayer interaction in ultrathin Cu (001)/Ni/Fe multilayers, and achieved a field-free skyrmions phase at room temperature (see Fig. 4(d)).[79] The FM/HM multilayer systems are of high research interest as skyrmions hosting materials, with the common strategies to use two FM/HM interfaces with opposite iDMI chirality, and to use both the FM/oxide and FM/HM interfaces (see Figs. 4(e)-(g)).[80-85]

Due to the non-trivial topology of skyrmions, charge carriers can feel an extra force while they pass skyrmions.[86-88] It has also been remarked that, compared to the current-driven DW motion, the critical current required for SOT- and STT-driven skyrmion motion can be much lower.[89-93] However, skyrmions are deflected sideways when moving, as a manifestation of their non-trivial topology which is called gyrotropic force, or skyrmion Hall effect (SkHE).[94-96] This gives rise to a transverse velocity during the current driven skyrmion dynamics (see Fig. 4(h)). With synthetic antiferromagnetic (SAF) structures (see Fig. 4(i)), the gyrotropic force in the upper and lower magnetic layers compensate each other, so that skyrmions can be

driven in a straight racetrack.[97]

# 7. Applications

The DMI is one of the important spin-orbit properties at the basis of spin-orbitronics and its applications. For example, DMI is essential in the concept of the chiral Néel DWs, which are involved in the current developments of racetrack memories, and plays an important role in the switching of SOT-RAMs for logic and memory functions.[36] The concept of racetrack memory introduced by Parkin in 2008 was based on motion of DWs driven by STT in magnetic films with in-plane magnetization, see Fig. 5(a).[98] The situation changed with the demonstration of the stabilization of chiral Néel DWs in perpendicularly magnetized magnetic films and the prediction of their fast motion by SOT.[36] As the SOT-induced fast motion of such Néel DWs was rapidly confirmed by Emori *et al.* and Kwang-Su Ryu *et al.*, the most recent efforts for the development of DW-based racetrack memory have been performed in this direction.[50,51] In 2015, Yang and Parkin proposed a racetrack memory based on the chiral Néel DWs in SAF structures, which can further increase memory speed and minimize the size of devices.[99] In the recent years, the research on magnetic devices has considered exploiting the motion of skyrmions, as described below.

With skyrmions, a variety of devices have been proposed including storage, logic and neuromorphic devices. Skyrmion racetrack memories based on HM/FM films and current-induced motion of skyrmions were first proposed by Fert *et al.*,[90] in which "0" and "1" states are associated to the absence or presence of one skyrmion, see Fig. 5(b). As the spacing between neighboring skyrmions can be of the order of their diameter or, approximately, of the order of a DW width, one can expect a higher memory density with skyrmions than with DWs. By placing two magnetic tunnel junctions (MTJ) on a racetrack of HM/FM film to generate and to detect skyrmions states, Zhang *et al.* proposed the magnetic skyrmion transistor device through voltage-gate control.[100,101] For FM/oxides interface, DMI could be modulated by ion-gating.[102] With the ion gating technique, Fillion *et al.* realized reversible control of skyrmions chirality in FeCoB/TaO$_x$ multilayers.[103] From such designs, spin-logic devices based on skyrmions have been intensively studied.[104-107] As skyrmions are encodable particle-like structures, multiple skyrmions could also be used as a multi-valued memory.[108-110] This type of concept is also suitable for neural network related applications.[111]

Apart from electric current, skyrmions can also be driven by electrical field, temperature gradient, and spin waves, which could also enable potential applications.[112-115]

Finally, DMI plays an important role in several other applications of spin-orbitronics. For example, spin tilts by DMI on the sample's edges, and motions of DMI-induced Néel DW are involved in the switching of perpendicular magnetization by SOT in devices of SOT-MRAM type.[116,117] Recently, Yu *et al.* show that perpendicular magnetization switching can be realized by DMI torque.[118]

## 8. Perspective

Until now, most of our discussion was focused on the FM-based multilayer films, with inversion symmetry along the axis perpendicular to the films broken by an interface, thus making the presence of iDMI inevitable. Some of the FM-nonmagnetic (NM) -FM stacks could also break the in-plane inversion symmetry, and accordingly lead to an interlayer DMI coupling the spins in successive layers.[119-122] In the simplest situation, the interlayer DMI can be expressed as a coupling between the magnetizations $\boldsymbol{m_1}$ and $\boldsymbol{m_2}$ of the top and bottom layers, $E_{DMI} = -\boldsymbol{D_{1,2}} \cdot (\boldsymbol{m_1} \times \boldsymbol{m_2})$. This coupling leads to a small canting of the magnetizations $\boldsymbol{m_1}$ and $\boldsymbol{m_2}$ from the perpendicular direction, as represented in Fig.6(e), and to the possibility of field free switching by SOT.[123]

Moreover, composition gradient and oblique growth of ultrathin films can also lead to additional symmetry breaking in multilayers, resulting in a gradient-induced DMI (g-DMI), that inevitably comprises bulk DMI components.[122,124-128] The presence of g-DMI can facilitate perpendicular magnetization switching in SOT devices.[129,130]

Recent breakthroughs in realizing two-dimensional (2D) intrinsic ferromagnetic films offer other candidates for future spintronics.[131-134] However, except for some rare cases, most of the 2D magnets obtained from exfoliation are centrosymmetric, resulting in a vanished total DMI.[33] Experimental and theoretical reports demonstrate that fabricating van der Waals heterostructures (see Fig. 6(a)) and use of chemical absorption (see Fig. 6(b)) can be effective to introduce ISB in the 2D magnets, and thereby generate sizable DMI to stabilize skyrmionic spin textures.[135-137] Another strategy to introduce ISB in the 2D magnets is to artificially build "Janus" magnets such as MnSeTe, CrGe(Si,Te)$_3$ , CrSeTe, and etc.[138-145] As shown in Fig.6 (c), the DMI values in Janus magnets MnSeTe and MnSTe are strong enough to stabilize skyrmions. More recently, researches showed that antiskyrmions could be found in a group of 2D magnets with P-4m2 space group (see Fig. 6(d)).[146,147]

Fascinating 2D magnets belong to the type-I 2D multiferroics, in which the coupling between magnetism and electric polarization (ME) could provide a convenient way of electric field control of magnetism. Theoretical models predicted that DMI chirality and strength in magnets depend strongly on the electric polarization.[148,149] Since the 2D multiferroics such as the $VOI_2$ monolayer and the $Ca_3FeOsO_7$ bilayer harbor intrinsic ISB, DMI and topological magnetic textures tuned by an electrical field can be realized.[150,151] Furthermore, in the CrN monolayer (see Fig. 6(f)) and the Co $(MoS_2)_2$ monolayer, the transformation between four states of skyrmions can be tailored by an out-of-plane electric field.[31,152]

In addition, it is inevitable to induce ripples for a 2D material either freestanding or on a substrate as soon as the size of 2D material is large enough. If such a curved system is magnetic, it is highly possible to achieve DMI in low dimensional magnets.[153] The presence of DMI in curved one-dimensional $CrBr_2$ and 2D $MnSe_2$,[154] as well as $CrI_3$ nanotubes has been theoretically confirmed.[155] Lastly, the twisting technique can also introduce ISB for moiré lattice 2D materials.[156,157] Therefore, DMI and skyrmions could be induced in moiré lattice 2D magnets.[157, 158]

**ACKNOWLEDGMENTS** This work was supported by the National Key Research and Development Program of China (MOST) (Grants Nos. 2022YFA1405100 and 2022YFA1403601), the National Natural Science Foundation of China (Grant No. 12174405), and the European Union's Horizon 2020 research and innovation Program under grant agreement 881603 (Graphene Flagship).

**Figure captions:**

**FIG. 1.** (Color online) Schematic representations of (a) spin structures of three states of $Fe_2O_3$ proposed by I.Dzyaloshinskii.[4,5] (b) the Moriya rules.[8] (c) chirality of DMI (left panel); DMI model proposed by A. Fert and P. M. Lévy.[12,13]

**FIG. 2.** (Color online) Schematic representations of (a) non-SOC spin spirals; (b) SOC included homogenous plane spin spiral. Calculation of DMI based on first-order perturbation of generalized Bloch theroem for (c) ferromagnetic Rh (111)/Fe/Pd system, adapted from Ref. 28 (©2018 American Physical Society); (d) antiferromagnetic Rh (001) /Ir/Fe system.[28,29] Adapted from Ref. 29 (©2017 American Physical Society).

**FIG. 3.** (Color online) (a) Schematic representations of a Bloch wall and chiral Néel domain walls with opposite chiralities. (b) Chiral Néel domain walls observed by Chen.[39] Adapted from Ref. 39 (©2013 Springer Nature). (c) Schematic representations of STT and SOT in FM/HM heterostructures (upper panels), and of the SOT effect on Bloch and chiral Néel walls (lower panels). (d) Total energy difference calculation of iDMI for Co(3ML)/Pt(3ML), where $d^k$ denotes DMI energy from each atomic layer k, and $\Delta E_{SOC}^{kk\prime}$ is the SOC energy associated to DMI from each atomic layer.[54] Adapted from Ref. 54 (©2015 American Physical Society).

**FIG. 4.** (Color online) First observed skyrmions in (a) B20 FeCoSi bulk material, adapted from Ref. 70 (©2010 Springer Nature); (b) Ir(111)/Fe ultrathin film structure.[69-71] Adapted from Ref. 70 (©2010 Springer Nature). (c) Spin textures of Bloch type skyrmion, Néel type skyrmion and antiskyrmion. (d) R.T. skyrmions discovered in the Cu/Ni/Fe multilayer heterostructures.[79] Adapted from Ref. 79 (©2015 American Institute of Physics). (e) Schematic representation of DMI in Pt/Co/Ir mulitlayers. Observed skyrmions in Pt/Co/Ir multilayers (f), adapted from Ref. 80 (©2016 Springer Nature); and (g) adopted from Ref. 81 (©2016 Springer Nature).[80,81] Schematic representation of (h) Skyrmion Hall effect,[93] adopte frome Ref. 93 (© 2013 Nature Publishing Group); (i) Skyrmions in synthetic antiferromagnetic bilayers,[97] from Ref. 97 (©2016 The authors).

**FIG. 5.** (Color online) (a) Domain wall racetrack memory.[98] (b) Skyrmions based racetrack devices.[90]

**FIG. 6.** (Color online) (a) Skyrmion in a $WTe_2/Fe_3GeTe_2$ van der Waals structure.[134] from Ref. 134 and licensed under CC-BY-4.0 (©2020 The Author(s)). (b) Skyrmions in oxidized $Fe_3GeTe_2$ few layer structures.[137] from Ref. 137 (©2021 American Physical Society). (c) DMI and skyrmions in Janus Mn$XY$ ($X,Y$ = S,Se and Te) monolayers.[138] Adapted from Ref. 138 (©2020 American Physical Society). (d) Anisotropic DMI and topological spin textures in 2D magnet with P$\bar{4}$m2 space group,[146] from Ref. 146 (©2020 American Chemistry Society). (e) Left: Sputtering of perpendicular synthetic antiferromagnets CoFeB/Pt/Ru/Pt/CoFeB with Ru for AF coupling and tilted sputtering of the bottom CoFeB layer to induce in-plane asymmetry and interlayer DMI. Right: Resulting interlayer DMI (see vector $D_{12}$) and opposite tilts of the magnetization in top and bottom layers.[123] Adapted from Ref. 123 and licensed under CC-BY-4.0 (©2023 The Author(s)). (f) Four states of skyrmions in the 2D multiferroic magnet CrN with chirality (A, C) controlled by electric field and polarity (up, down) by magnetic field.[31] Adapted from Ref. 31 (©2020 American Physical Society).

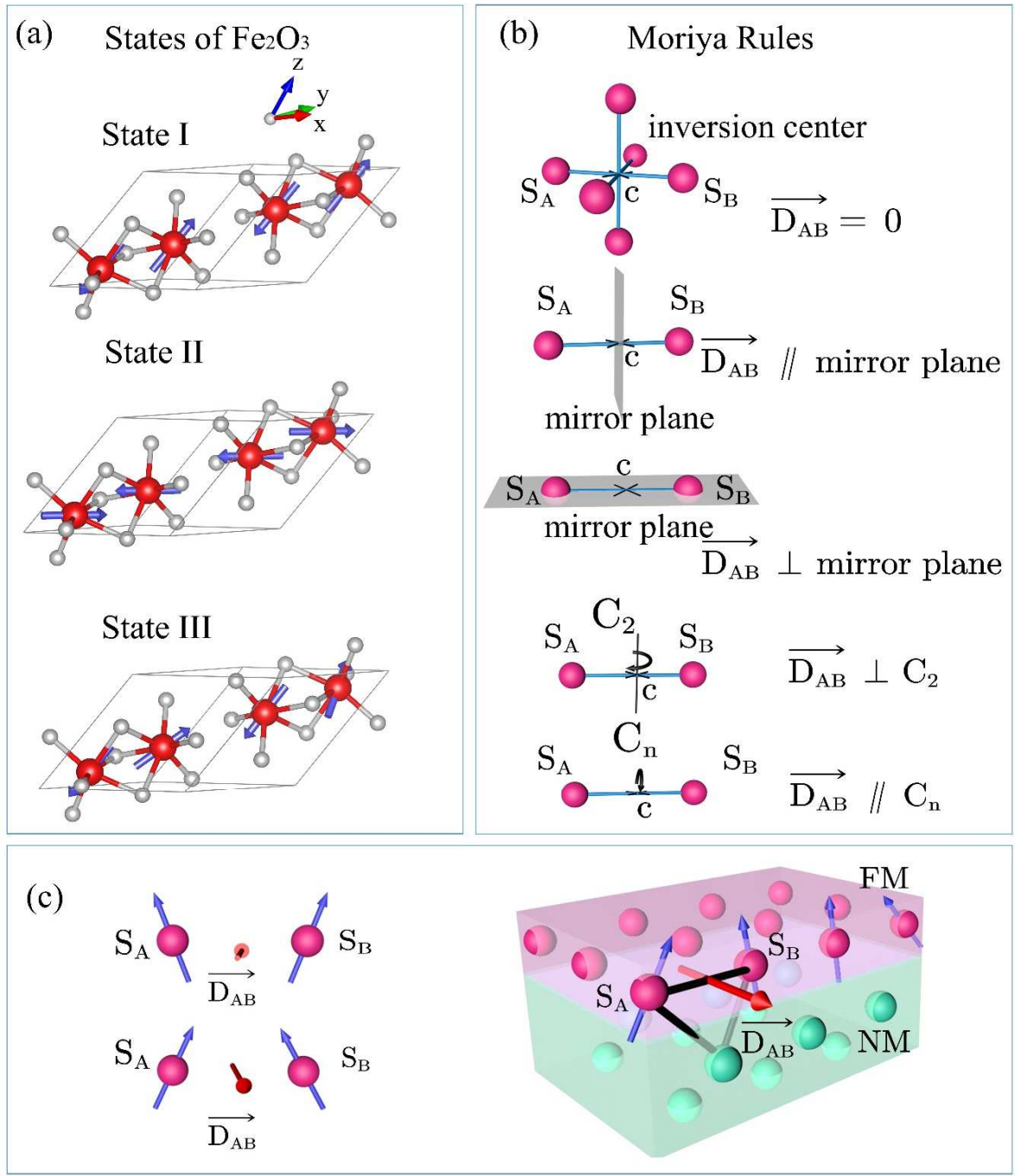

**FIG. 1.** (Color online) Schematic representations of (a) spin structures of three states of $Fe_2O_3$ proposed by I.Dzyaloshinskii.[4,5] (b) the Moriya rules.[8] (c) chirality of DMI (left panel); DMI model proposed by A. Fert and P. M. Lévy.[12,13]

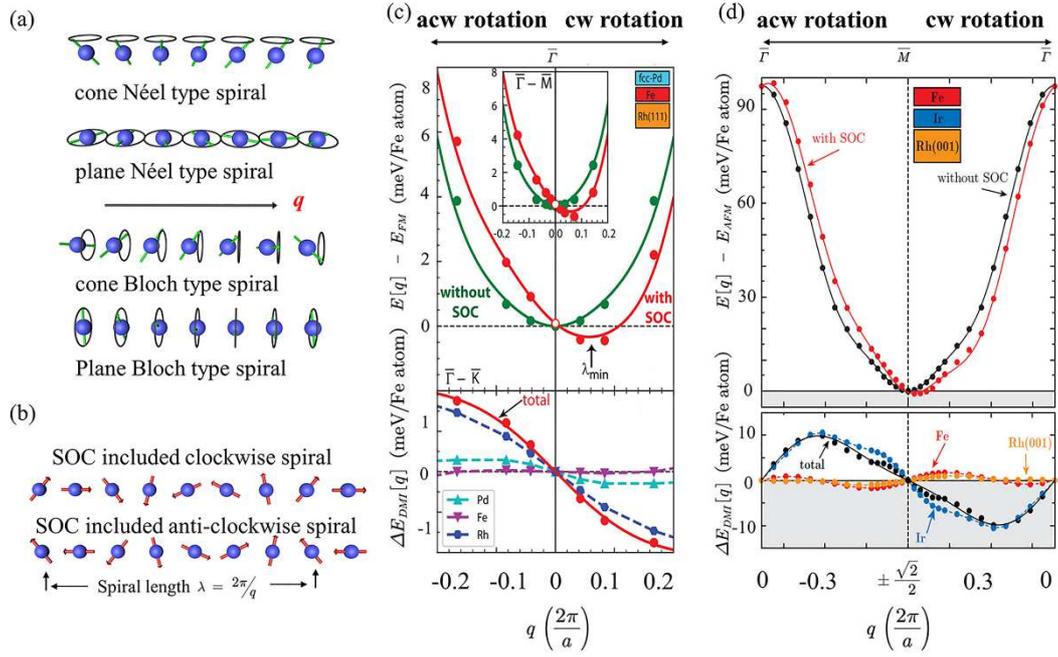

**FIG. 2.** (Color online) Schematic representations of (a) non-SOC spin spirals; (b) SOC included homogenous plane spin spiral. Calculation of DMI based on first-order perturbation of generalized Bloch theroem for (c) ferromagnetic Rh (111)/Fe/Pd system, adapted from Ref. 28 (©2018 American Physical Society); (d) antiferromagnetic Rh (001) /Ir/Fe system.[28,29] Adapted from Ref. 29 (©2017 American Physical Society).

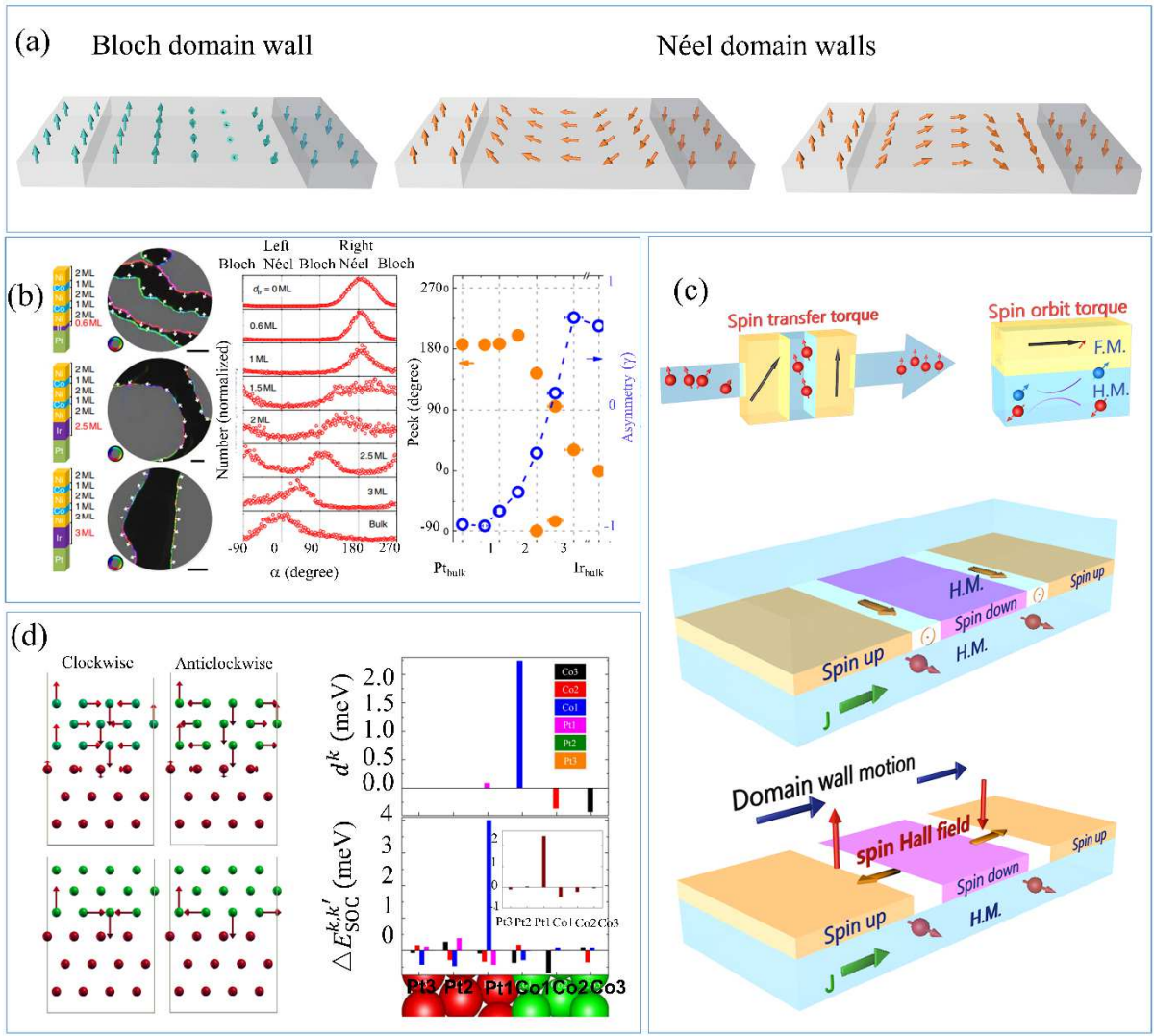

**FIG. 3.** (Color online) (a) Schematic representations of a Bloch wall and chiral Néel domain walls with opposite chiralities. (b) Chiral Néel domain walls observed by Chen.39) Adapted from Ref. 39 (©2013 Springer Nature). (c) Schematic representations of STT and SOT in FM/HM heterostructures (upper panels), and of the SOT effect on Bloch and chiral Néel walls (lower panels). (d) Total energy difference calculation of iDMI for Co(3ML)/Pt(3ML), where $d^k$ denotes DMI energy from each atomic layer k, and $\Delta E_{SOC}^{kk\prime}$ is the SOC energy associated to DMI from each atomic layer.54) Adapted from Ref. 54 (©2015 American Physical Society).

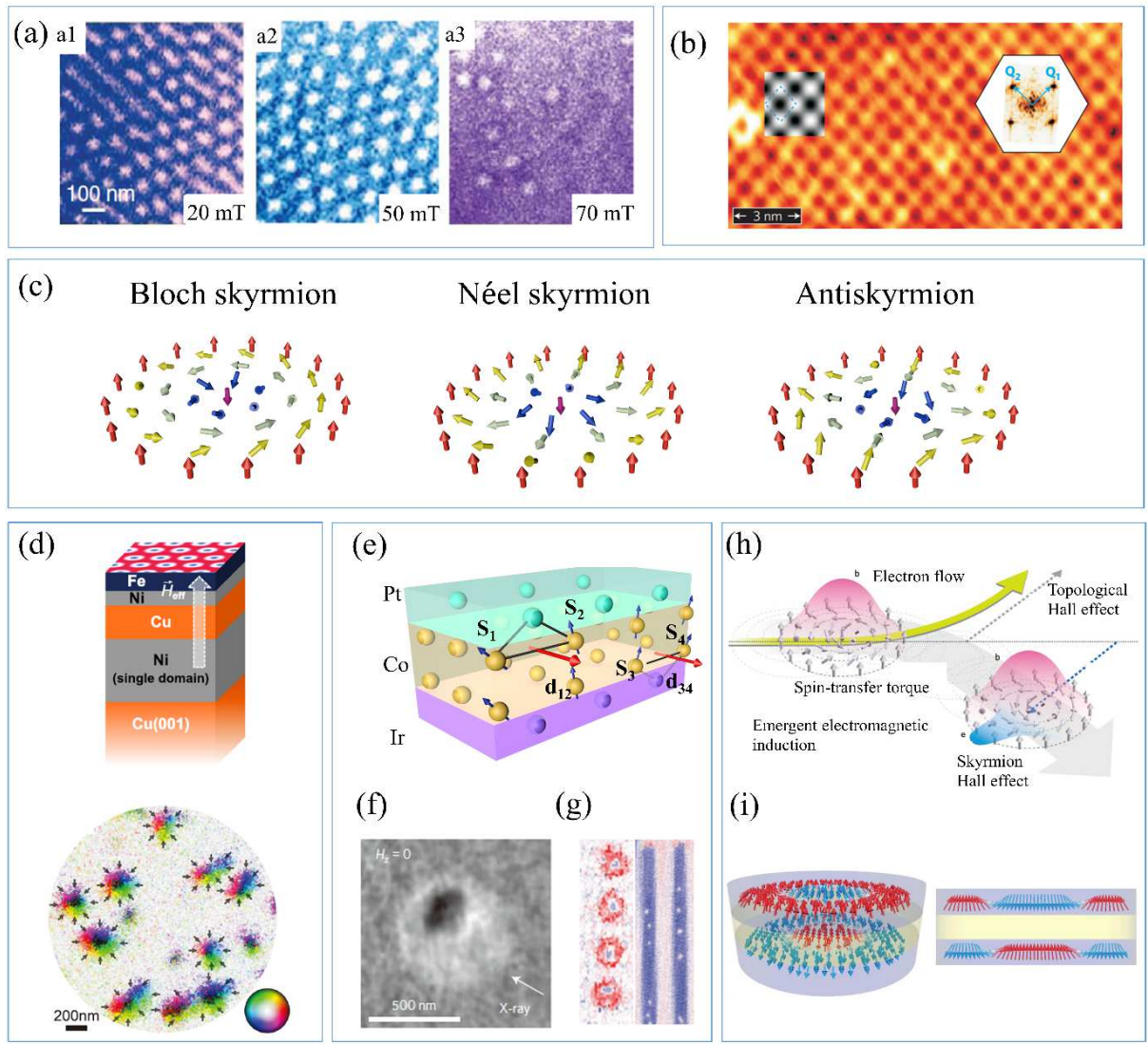

**FIG. 4.** (Color online) First observed skyrmions in (a) B20 FeCoSi bulk material, adapted from Ref. 70 (©2010 Springer Nature); (b) Ir(111)/Fe ultrathin film structure. [69-71] Adapted from Ref. 70 (©2010 Springer Nature). (c) Spin textures of Bloch type skyrmion, Néel type skyrmion and antiskyrmion. (d) R.T. skyrmions discovered in the Cu/Ni/Fe multilayer heterostructures.[79] Adapted from Ref. 79 (©2015 American Institute of Physics). (e) Schematic representation of DMI in Pt/Co/Ir mulitlayers. Observed skyrmions in Pt/Co/Ir multilayers (f), adapted from Ref. 80 (©2016 Springer Nature); and (g) adopted from Ref. 81 (©2016 Springer Nature).[80,81] Schematic representation of (h) Skyrmion Hall effect, [93] adopte frome Ref. 93 (© 2013 Nature Publishing Group); (i) Skyrmions in synthetic antiferromagnetic bilayers, [97] from Ref. 97 (©2016 The authors).

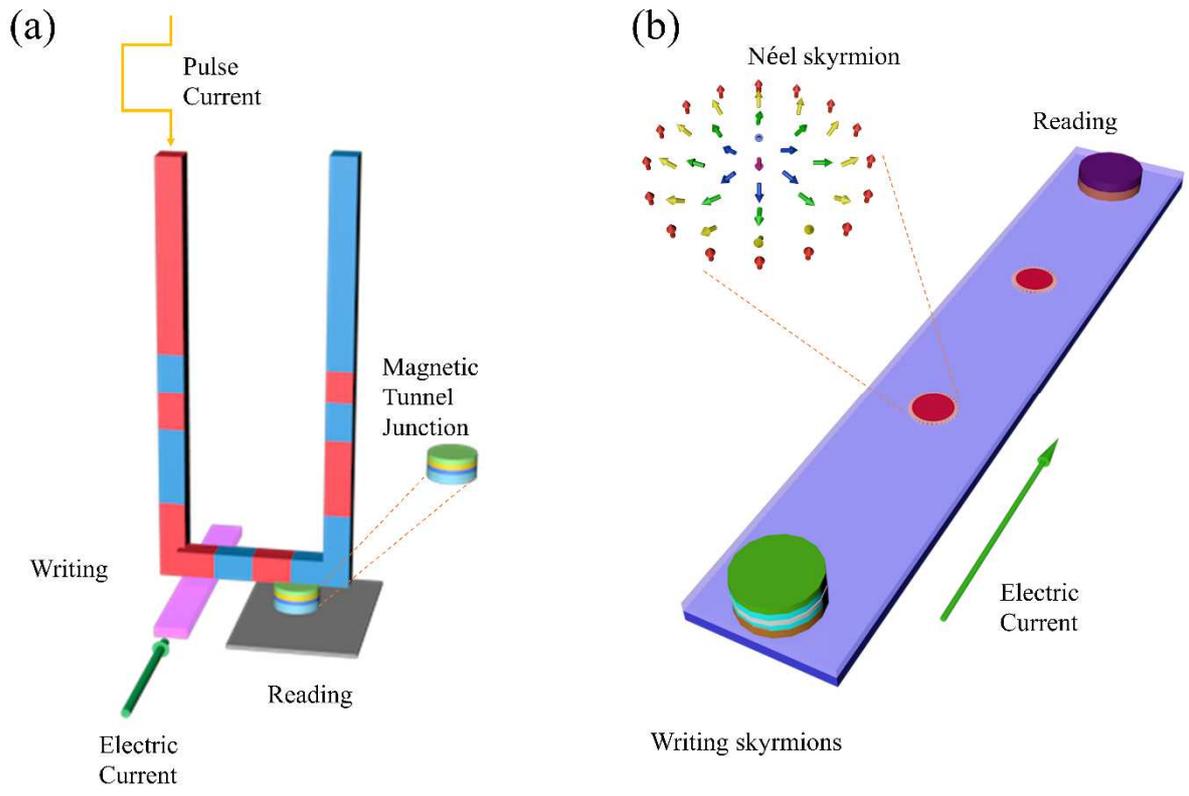

**FIG. 5.** (Color online) (a) Domain wall racetrack memory.[98)] (b) Skyrmions based racetrack devices.[90)]

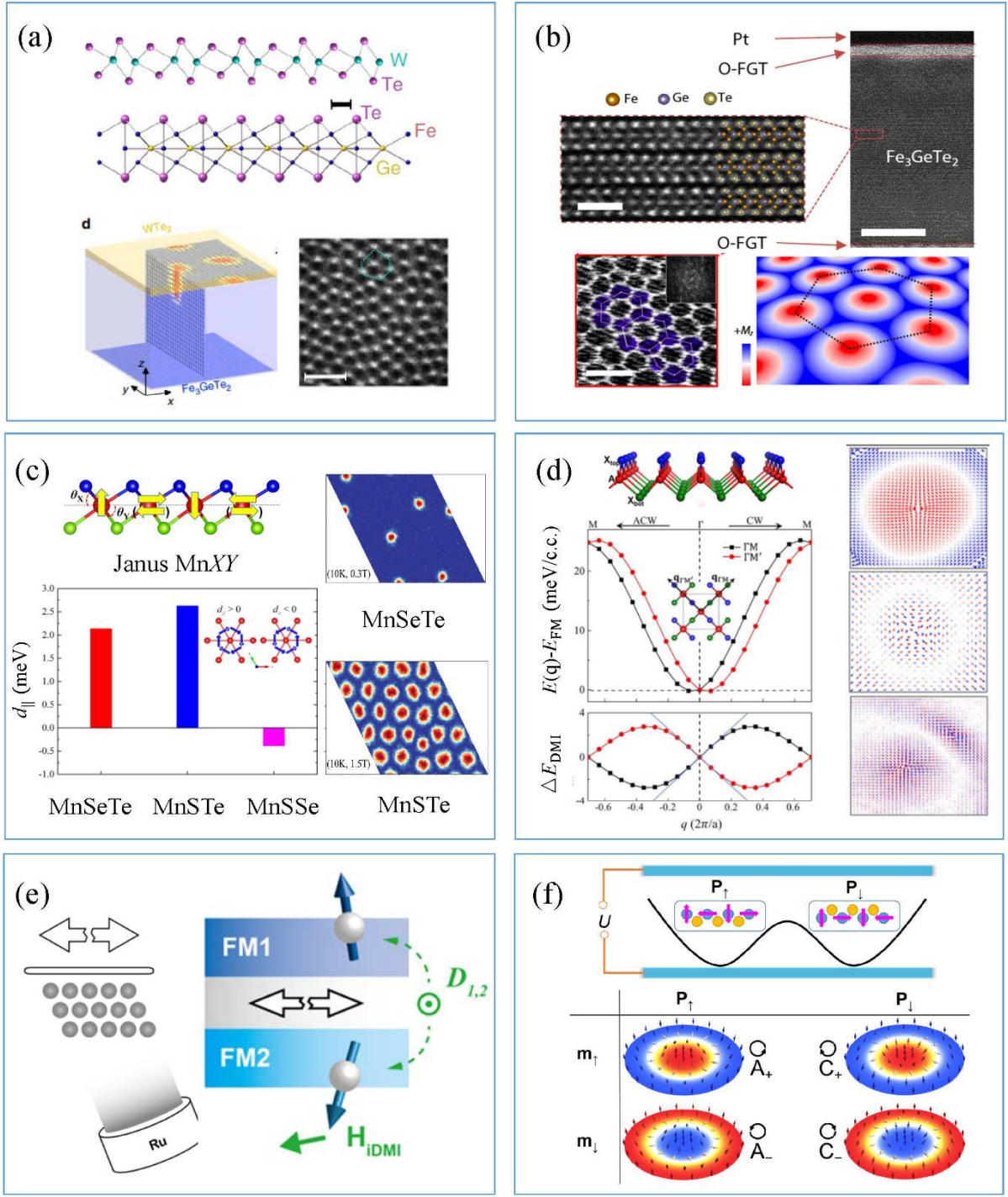

**FIG. 6.** (Color online) (a) Skyrmion in a WTe$_2$/Fe$_3$GeTe$_2$ van der Waals structure.[134] from Ref. 134 and licensed under CC-BY-4.0 (©2020 The Author(s)). (b) Skyrmions in oxidized Fe$_3$GeTe$_2$ few layer structures.[137] from Ref. 137 (©2021 American Physical Society). (c) DMI and skyrmions in Janus Mn$XY$ ($X,Y$ = S,Se and Te) monolayers.[138] Adapted from Ref. 138 (©2020 American Physical Society). (d) Anisotropic DMI and topological spin textures in 2D magnet with P$\bar{4}$m2 space group,[146] from Ref. 146 (©2020 American Chemistry Society). (e) Left: Sputtering of perpendicular synthetic antiferromagnets CoFeB/Pt/Ru/Pt/CoFeB with Ru for AF coupling and tilted sputtering of the bottom CoFeB layer to induce in-plane asymmetry and interlayer DMI. Right: Resulting interlayer DMI (see vector D$_{12}$) and opposite tilts of the magnetization in top and bottom layers.[123] Adapted from Ref. 123

and licensed under CC-BY-4.0 (©2023 The Author(s)). (f) Four states of skyrmions in the 2D multiferroic magnet CrN with chirality (A, C) controlled by electric field and polarity (up, down) by magnetic field.[31] Adapted from Ref. 31 (©2020 American Physical Society).

Authors information:

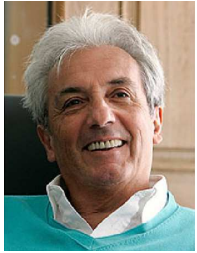

**Albert Fert** He is one of the discoverers of giant magnetoresistance which brought about a break-through in gigabyte hard disks and led to the development of spintronics. Currently, he is an emeritus professor at Paris-Saclay University, scien- tific director of a joint laboratory (Unité mixte de recherche) between the Centre national de la recherche scientifique (National Scientific Research Centre) and Thales Group. He was awarded the 2007 Nobel Prize in Physics together with Peter Grünberg.

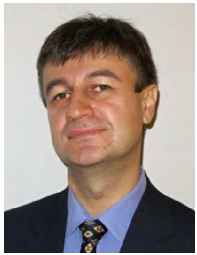

**Mairbek Chshiev** He is a theoretical physicist specializing on theory of spintronic phenomena in magnetic nanostructures and electronic structure of materials for spintronics using condensed matter theory and computational material science ap- proaches. He is the Head of Theory Group at SPINTEC in Grenoble where he moved from the U. S. as a holder of Chair of Excellence position at Nanosciences Foundation (2008–2011). He has been an invited researcher at University of Lorraine (2007), CNRS/Thales (2011) as well as adjunct associate professor at University of Alabama (2008–2010). He is an IEEE Senior Member and a member of the American (APS) and European (EPS) Physical Societies. In 2020 he was appointed Senior Member of the Institut Universitaire de France (IUF) and serves also as Director of European School on Nanosciences & Nanotechnologies (ESONN).

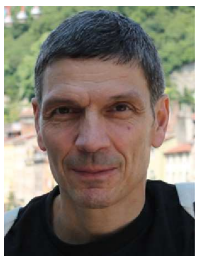

**André Thiaville** He is research director at the National Center for Scientific Research (CNRS) at the Laboratory of Solid State Physics at the Université Paris-Saclay in Orsay. He is a specialist of the structure and dynamics of magnetic objects such as domain walls, Bloch lines, Bloch points, Néel lines, magnetic vortices. His more recent work has focussed on the motion of magnetic walls induced by a spin-polarized electric current, be it by spin-transfer torque or by spin-orbit torques, where in particular he uncovered the key role of the interfacial Dzyaloshinskii-Moriya interaction. He has been a guest researcher in 1991–1992 at ETL, Tsukuba (with Dr. Y. Suzuki) and in 2006 at the ICR, Uji (with Pr. T. Ono). He is a member of IEEE France section, the Magnetic Society of Japan and the Société française de physique (French Physical Society).

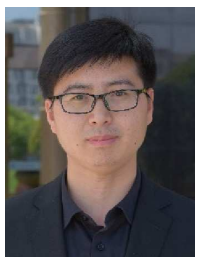

**Hongxin Yang** Professor at School of Physics in Nanjing University. He graduated in Physics from Jilin University in China, then received a PhD from University of Grenoble in France. After that, he did two stages of postdoctoral fellows at the Laboratory solid state physics at Paris-Sud University and the joint laboratory (Unité mixte de recherche) between the Centre national de la recherche scientifique (National Scientific Research Centre) and Thales Group in France. In 2017, he got a professor position in Chinese Academy of Sciences, and in 2022 he joined Nanjing University. He is a theoretical physicist in the field of first-principles calculations and micromagnetic simulations for spintronics. The first-principles calculation for DMI using real space spin-spiral method they developed has been one of the most popular methods in use. He has published more than 100 papers, including Nat. Rev. Phys., Nat. Mater., Nat. Nanotech., PRL., etc. He also gave many invited talks in APS March Meeting and InterMag etc. He became an IEEE senior member since 2022.